\documentclass[letterpaper,12pt]{article}   %% LaTeX 2e (preferred)
\usepackage{osajnl2} %% do not use with REVTeX4
\usepackage[draft]{hyperref} %% optional
\usepackage{fancyhdr, color, amssymb, amsmath}
\usepackage{graphicx}
\usepackage{subfig}
\usepackage{cite}
\usepackage[utf8]{inputenc}

\newcommand{\im}[0]{\text{Im}\,}
\newcommand{\re}[0]{\text{Re}\,}
\newcommand{\diff}{\text{d}}

\begin{document}

\title{The effect of gain saturation in a gain compensated perfect lens}

%% For REVTeX it is possible to automate superscript and e-mail callouts with the superscriptaddress option; see REVTeX4 documentation.

\author{Marte P. Hatlo Andresen,$^{1,*}$ Aleksander V. Skaldebø,$^{2,3}$ Magnus W. Haakestad$^4$, Harald E. Krogstad$^1$, and Johannes Skaar$^{2,3}$}
\address{$^1$Department of Mathematical Sciences, Norwegian University of
  Science and Technology, \\ NO-7491 Trondheim, Norway}
\address{$^2$Department of Electronics and Telecommunications, Norwegian University of Science and
Technology, \\ NO-7491 Trondheim, Norway}
\address{$^3$University Graduate Center, \\ NO-2027 Kjeller, Norway}
\address{$^4$Norwegian Defence Research Establishment (FFI), P O Box 25, \\ NO-2027 Kjeller, Norway}
\address{$^*$Corresponding author: hatlo@math.ntnu.no}

\begin{abstract} The transmission of evanescent waves in a gain-compensated perfect lens is discussed. In particular,
the impact of gain saturation is included in the analysis, and a method for calculating the fields
of such nonlinear systems is developed. Gain compensation clearly improves the resolution; however,
a number of nonideal effects arise as a result of gain saturation. The resolution associated with
the lens is strongly dependent on the saturation constant of the active medium.
\end{abstract}

% REPLACE WITH CORRECT OCIS CODES FOR YOUR ARTICLE
                          % NOTE: \ocis{} IS ALIASED TO \pacs{} BUT MUST
                          % FORMAT THE TERMS CORRECTLY FOR EACH JOURNAL

\maketitle %% null function with osajnl.sty

\section{Introduction}
Metamaterials have large potential in electromagnetics and optics due to their possibility of
tailoring the permittivity and permeability. This enables construction of, for example, media with
negative refractive index $n$ \cite{veselago}, perfect lenses \cite{pendry2000}, invisibility cloaks
\cite{pendry2006,leonhardt2006s}, and other exciting components transforming the electromagnetic
field \cite{leonhardt2006}.
Unfortunately, the performance of such devices is strongly limited by losses. Although causality and
passivity do not prohibit negative index materials with arbitrary low losses \cite{nistad08}, in
practice it is difficult to fabricate materials with high figure of merit $\text{FOM}=-\re n/\im n$,
 especially at optical frequencies \cite{shalaev_nphot,shalaev05,dolling,kwon}. For perfect lenses,
losses limit the amplification of evanescent waves associated with large spatial frequencies, and
therefore the resolution \cite{ramakrishna2002,johansen}. It has therefore been suggested to
introduce gain into the metamaterials
\cite{shalaev_nphot,ramakrishna_prb,noginov06,popov06,klar2006,zhang2009,fang2009,
sivan2009, vincenti,dong}.
This could be a promising approach provided the intrinsic losses can be made relatively small so
that compensation by a realistic amount of gain is possible.

Both permittivity $\epsilon$ and permeability $\mu$ may involve losses; thus in general, gain may be
needed to reduce \emph{both} $\im\epsilon$ and $\im\mu$.
For a perfect lens it is generally not sufficient e.g. to reduce $\im\epsilon$ below zero such that
the refractive index $n=\sqrt{\epsilon\mu}$ becomes real. However, as long as the object to be
imaged is one-dimensional, only one polarization (TE or TM) of the electromagnetic field is
required. Then, provided the lens is sufficiently thin, only one of the parameters $\epsilon$ and
$\mu$ is relevant for the transmission of evanescent waves \cite{pendry2000,johansen}. Choosing TM
polarization, only $\epsilon$ matters, enabling gain compensation with dielectric, active media.

Introducing the necessary active material into a metamaterial leads to a change not only in the
imaginary part but also the real part of the permittivity, and should therefore be kept in mind
while designing the metamaterial structure.
Other critical considerations include matching of the negative refractive index frequency band to
that of the gain lineshape function, the level of loss possible to overcome in the absence of
saturation, and the saturation constant of the active medium.

There have been several attempts to create a perfect lens in the near~IR-spectrum the last years
\cite{shalaev_nphot,shalaev05,dolling,enkrich}. The FOM currently reported is of the order of 3 for
the frequency where $\re n\approx-1$ \cite{shalaev05,dolling}.
With these values, traditional optical amplifiers such as Erbium-doped silica or gas laser
amplifiers will not be able to reduce the intrinsic losses significantly.
Theoretical studies have shown that it may be possible to raise the FOM at near~IR-frequencies to as
much as 20, while keeping $\re n\approx -1$ \cite{zhang}.
It has also been reported that laser dyes, or dye-Ag aggregate mixtures, may reach amplifications of
up to $\im n\approx-0.06$ at near~IR-frequencies \cite{noginov06,lawandy}.
Taking into account these reports, this article will not speculate further on the choice and design
of the metamaterial, but merely assume an appropriate material is physically feasible.

The main purpose of our work is to consider the transmission of evanescent waves in a practical,
gain-compensated perfect lens. Clearly, gain saturation is highly relevant in this context, and we
demonstrate how this effect leads to limited amplification of evanescent fields, and therefore
limited resolution. We calculate the resolution as a function of the saturation constant of the
active medium, and also the detailed field profile and reflections
from the lens. It will become clear that gain saturation is a critical effect which
may lead to severe limitations.

\section{Nonlinear gain saturation and field calculations}

The relative permittivity of the active metamaterial is given by:
\begin{equation}\label{eq:permittivity}
\epsilon(\omega) = 1 + \chi_\text{p}(\omega) + \chi_\text{a}(\omega),
\end{equation}
where $\chi_\text{p}(\omega)$ denotes the susceptibility of the passive structure, and
$\chi_\text{a}(\omega)$ the contribution from the active part. The contribution to the
susceptibility for travelling and evanescent fields in active media can be modeled using
semiclassical theory \cite{scully}. If there is spherical
symmetry, allowing for coupling to several degenerate states with different values of the quantum
number $m$, one can show using the Wigner-Eckart theorem and the general properties of the
Clebsch-Gordan coefficients that the system can effectively be treated
as a two-level system. Assuming two-level atoms and
using the dipole approximation, we find the following expression for the active susceptibility:
\begin{equation}\label{eq:chia}
\chi_\text{a}(\omega) = 
\frac{A(\omega)\left( \frac{\omega - \omega_0}{\gamma} - i \right)}{1 +
\frac{\left|\mathbf{E}\right|^2}{\mathcal{E}_\text{s}(\omega)^2}}.
\end{equation}
Here $\omega_0$ is the transition frequency, $\omega$ the frequency of the incident light, $\gamma$
a phenomenological decay rate due to spontaneous emission and 
elastic collisions, $\mathbf{E}$ the complex electric field, and $\mathcal{E}_\text{s}(\omega)$ the
saturation constant of the active medium. The saturation constant depends on the selected gain
material and pumping level. For dye amplifiers a normal value is
$\mathcal{E}_\text{s}(\omega_0)\sim10^7~$V$/$m \cite{lawandy,destro}, corresponding to intensities
in the $\text{kW}/\text{cm}^2$ regime for propagating waves. The numerator in Eq. \eqref{eq:chia}
describes the susceptibility in the limit  $|\mathbf{E}|/\mathcal{E}_s
\rightarrow 0$. 
The numerator contains the line shape function and several material parameters;
 factors that are irrelevant for the analysis below are absorbed into the function $A(\omega)$. For $\omega=\omega_0$ and
 $|\mathbf{E}|/\mathcal{E}_s
\rightarrow 0$, $A(\omega)$ is simply $-\im\chi_\text{a}(\omega)$. Both functions $A(\omega)$
and $\mathcal{E}_\text{s}(\omega)$ are real-valued.
Note that the material parameters describing the active medium are
effective parameters, depending on the geometry of the metamaterial
structure, and are not necessarily equal to the bulk parameters of the gain
material \cite{fang2009,
sivan2009}. 

Throughout this paper, we will consider the frequency $\omega=\omega_0$. Eq. \eqref{eq:chia}
now reduces to:
\begin{equation}\label{eq:chia0}
\chi_\text{a}(\omega_0) = 
\frac{- iA(\omega_0)}{1 + \frac{\left|\mathbf{E}\right|^2}{\mathcal{E}_\text{s}(\omega_0)^2}}.
\end{equation}
Eq. \eqref{eq:chia0} describes how the imaginary part of the total
permittivity relates to the pumping and local field amplitude. %Note that
The real part of the total permittivity is independent of the pumping and
the local field.

Note that there is a fundamental difference between the nonlinearity due to
gain saturation, and conventional second- and third-order nonlinearities. First,
the nonlinearity due to the denominator in Eq. \eqref{eq:chia} is so large that
a Taylor expansion up third order is generally not valid. Second, the
nonlinearity of a gain medium is expressed in terms of the
slowly varying field envelope $|\mathbf E|$, while the second- and third
order nonlinearities usually are expressed in terms of the rapidly varying
time-dependent electric field. Since the nonlinearity in our case can be
characterized using $|\mathbf E|$, the medium will not generate new
frequencies for monochromatic input 
\cite{svelto}. Nevertheless, if several modes (or frequencies) are
present, the modes interact in the sense that the complex refractive index seen
by one mode is dependent on the presence and strength of all modes.

\begin{figure}
\centering
\includegraphics[width=0.7\textwidth]{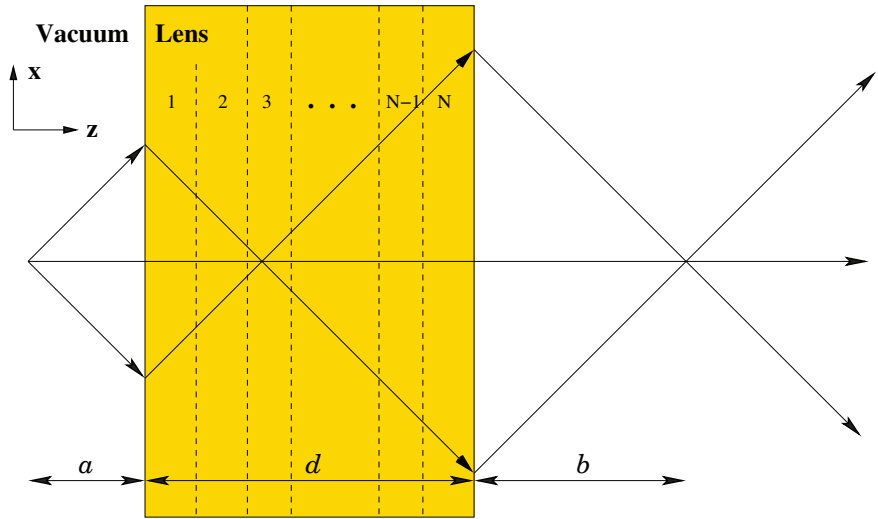}
\caption{
(Color online) Perfect lens in vacuum. The parameter $d$ is the thickness of the lens, $a$ and $b$ are the
distances from the source to the lens, and from the lens to the image
plane, respectively. The parameters are
governed by the equation $d = a + b$. The numbers $1$ through $N$ indicate the different slices. The
lens is considered to be infinite in the $xy$-plane. ASHKSF1.eps.
}
\label{fig:simple}
\end{figure}

We consider a perfect lens slab which extends to infinity in the $xy$-plane, and has thickness $d$
in the $z$-direction, see Fig. \ref{fig:simple}. The source is located a distance $a$ (with $a<d$)
from the input end of the lens. The incident field from the source will be taken to be a
superposition of plane TM-waves, with the magnetic field in the $y$-direction. Provided $\omega_0
d/c\ll 1$ and $|\mu|\sim 1$, the spesific value of $\mu$ is not critical for the operation of the
lens for evanescent waves \cite{pendry2000,johansen}. Here $c$ is the vacuum light velocity. The permittivity is given by
Eqs. \eqref{eq:permittivity}, \eqref{eq:chia0}, and
$\re\chi_\text{p}(\omega_0)=-2$. The remaining losses
after gain compensation (in the absence of saturation) is described by the parameter:
\begin{equation}
\Delta\chi=\im\chi_\text{p}(\omega_0)-A(\omega_0). 
\end{equation}

To find the steady state solution to Maxwells equations for our nonlinear medium, an iterative
approach can be used. In the zeroth iteration, the electric field is simply set to zero everywhere.
(Alternatively, the initial field could be set to infinity. This does not give any significant
difference in performance, in terms of the required number of iterations.) In the next iteration,
Eqs. \eqref{eq:permittivity} and \eqref{eq:chia0} are used to find an approximation of the
permittivity of the lens. Taking the incident magnetic field to be unity (normalized), we can 
compute the magnetic and electric fields everywhere. Now we may repeat the iteration; calculate a new approximation of the
permittivity from the field, computing the resulting field from this new structure, etc. The
iteration procedure has an inherent stability, as growing fields leads to less gain in the medium,
and vice versa.

Nevertheless, inaccuracies and even divergence may arise if the number of slices is too low, so that
the field no longer can be treated as constant in each slice. In the case $\Delta\chi=0$ the
convergence seems to be particularly sensitive to the number of slices. An alternative to increasing
the number of slices to a very high number, is to regularize the iterative approach as follows:
Rather than setting the permittivity to that resulting from the field in the previous iteration, it
can be set to a weighted mean of the permittivities as resulting from the last two iterations. In
our computations, the permittivity in iteration $i$ (for $i\geq 2$) was set to 0.5 times the
permittivity calculated by the field from iteration $i-1$, pluss 0.5 times that resulting from
iteration $i-2$. For $i=1$ the permittivity was calculated using the field from iteration $0$. This
resulted in convergence after $\sim 20-35$ iterations. The weight factor 0.5 is somewhat arbitrary;
other choices are possible but may require a larger $N$ or number of iterations to obtain
convergence.

If the root mean square deviation of three successive iterations are within a specified limit
($10^{-12}$ for the relative permittivity in our computations), and strictly decreasing, the results
are deemed converge. Note that when the fields of subsequent iterations coincide, we have a valid
solution to Maxwell's equations with constitute relation as implied by Eqs. \eqref{eq:permittivity} and
\eqref{eq:chia0}. 

In general, the fields in one iteration, and therefore the permittivity
in the next iteration, will be dependent on $x$ and $z$. Thus the
computation of the fields in the next iteration requires the solution
to Maxwell's equations in an inhomogeneous structure. Note that in
each iteration, the structure is linear; the nonlinearity 
of the structure enters through the iteration. 
For the
linear calculation, we employ a transfer matrix technique, considering
the different plane waves in the structure. The lens is divided into
$N$ slices in the $xy$-plane, as seen in Fig. \ref{fig:simple}. These
slices must be sufficiently thin, such that the permittivity inside
each slice is approximately uniform in the $z$-direction. For this
condition to be valid for the next iteration as well, the resulting
field from the present iteration must also be approximately
constant. This means that $k_{x}d/N\lesssim 1$ for the transverse
wavenumbers $k_x$ that contribute significantly to the fields.

The electric field can be found using the Ampere--Maxwell's law:
\begin{equation}\label{amperemaxwell}
\mathbf E(x,z)=\frac{1}{-i\omega\epsilon(x,z)\epsilon_0}\nabla\times\mathbf H(x,z).
\end{equation}
With periodic boundary conditions in the $x$-direction, 
the magnetic
field and the permittivity  can be expanded in discrete Fourier series
\begin{eqnarray}\label{h-field}
\mathbf H(x,z) = H(x,z)\mathbf{\hat y} &=& \sum_{m} h_m(z)\exp(ik_{xm}x)\mathbf{\hat y}, \\
\epsilon(x,z) &=&\sum_{m} e_m(z)\exp(ik_{xm}x), \label{permittivity} 
\end{eqnarray}
for some Fourier coefficients $h_m(z)$ and $e_m(z)$. Here $k_{xm}=2\pi
m/L$, $L$ is the computational domain, and $\mathbf{\hat y}$ is the unit vector in the $y$-direction.

From Maxwell's equations, we find that the magnetic field satisfies
\begin{equation}
\label{Master_eq}
\nabla^2 H + \epsilon \mu k^2 H - \frac{1}{\epsilon} \frac{\partial
  \epsilon}{\partial x} \frac{\partial H}{\partial x} = 0,
\end{equation}
where $k=\omega_0/c$. We express  $1/\epsilon$ and $(1/\epsilon)
\partial \epsilon/\partial x$ as Fourier series as follows
\begin{equation}
\frac{1}{\epsilon(x)}  = \sum_m Q_m \exp(i k_{xm}x),
\end{equation}
and
\begin{equation}
\frac{1}{\epsilon(x)} \frac{\partial \epsilon(x)}{\partial x} = \sum_m F_m \exp(i k_{xm} x).
\end{equation}
The Fourier coefficients $F_m$ are now given as
\begin{equation}
\label{fm}
F_m = \sum_{m'} Q_{m-m'} i k_{xm'} e_{m'}, 
\end{equation}
or as a matrix product
\begin{equation}
F = i  \mathbf{Q}\mathbf{k_x} \mathbf{e},
\end{equation}
where  $F = \{F_m\}_{m}$, $\mathbf{k_x} = \mathrm{diag}(k_{xm})$, $\mathbf{e} =
\{e_m\}_{m}$ and $\mathbf{Q}$ is a Toeplitz matrix with elements $\mathbf{Q}_{i,j}=  Q_{i-j}$.

Inserting the Fourier series into equation (\ref{Master_eq}), we obtain
\begin{equation}
\label{helm}
\frac{\mathrm{d}^2 h_m (z)}{\mathrm{d} z^2} - k_{xm}^2
h_m(z) + k^2 \mu \sum_{m'} \varepsilon_{m-m'} (z)h_{m'}(z) - \sum_{m'}
i  F_{m-m'} k_{xm'} h_{m'} =0,
\end{equation}
for each $m$. Let $\mathbf{h} = \{h_m\}_{m\in \mathbb{Z}}$ %and define 
and use $k_{zm}^2 = k^2 - k_{xm}^2$. We can write Eq. (\ref{helm}) 
as a matrix equation
\begin{equation}
\label{eq5}
\frac{\mathrm{d}^2 \mathbf{h} (z)}{\mathrm{d} z^2} + \left[\mathbf{k_z}^2  +
  \mathbf{V} \right] \mathbf{h}(z) = 0,
\end{equation}
where $\mathbf{k_z} = \mathrm{diag}(k_{zm})$, and $\mathbf{V}$ is the
 operator defined as
\begin{equation}
\label{Toep_op}
\mathbf{V} = -k^2 \mathbf{I} + k^2 \mu \mathbf{G} - \mathbf{F}.
\end{equation}
Here $\mathbf{G}$ and $\mathbf{F}$ are infinite dimensional Toeplitz matrices with elements $\mathbf{G}_{i,j} = e_{i-j}$ and $\mathbf{F}_{i,j}=i  F_{i-j}k_{xj}$.

Eq. (\ref{eq5}) may be decomposed into a first order system by
writing $\mathbf{h} = \mathbf{h^+} +
\mathbf{h^-}$, where  $\mathbf{h^+}= \{h_m^+\}_{m \in\mathbb{Z}}$ and $\mathbf{h^-}=
\{h_m^-\}_{m \in\mathbb{Z}}$. In fact, the equations
\begin{subequations}\label{decomposition}
\begin{eqnarray}
  \frac{\diff \mathbf{h^+}}{\diff z} &=& i\mathbf{k_z}\mathbf{h^+} +
  i (2 \mathbf{k_z})^{-1} \mathbf{V} (\mathbf{h^+}+\mathbf{h^-}),\label{decomp1}\\
  \frac{\diff \mathbf{h^-}}{\diff z} &=& -i\mathbf{k_z}\mathbf{h^-} -
  i (2\mathbf{k_z})^{-1} \mathbf{V}(\mathbf{h^+}+\mathbf{h^-}),\label{decomp2}
\end{eqnarray}
\end{subequations}
are seen to be equivalent to (\ref{eq5}) after  differentiation and summation.
This  decomposition is particularly convenient, since outside the lens
$\mathbf{V}=0$, and Eq. (\ref{decomposition}) has the simple solution $h_m^\pm(z) = \text{const}\cdot\exp(\pm i k_{zm} z)$. In other words, outside the lens, $h_m^+$ and $h_m^-$ are the forward and backward propagating waves, respectively.

By adding Eqs (\ref{decomp1}) and (\ref{decomp2}), we obtain
\begin{equation}
\label{diff_h}
\frac{\diff \mathbf{h}}{\diff z} = i \mathbf{k_z} (\mathbf{h^+} - \mathbf{h^-}),
\end{equation}
which is needed for $\nabla \times \mathbf{H}$.

By writing
\begin{equation}
\Psi   \mathbf{=}\left[
\begin{array}
[c]{c}%
\mathbf{h^+}\\
\mathbf{h^-}
\end{array}
\right]  , \quad 
\mathbf{C}    \mathbf{=}\left[
\begin{array}
[c]{cc}%
i \mathbf{k_z} + i (2 \mathbf{k_z})^{-1} \mathbf{V} & i (2
\mathbf{k_z})^{-1} \mathbf{V}\\
 - i (2 \mathbf{k_z})^{-1} \mathbf{V} &-i \mathbf{k_z}- i (2
\mathbf{k_z})^{-1} \mathbf{V}\\
\end{array}
\right]  , \label{c}
\end{equation}
Eq. (\ref{decomposition}) may be brought into matrix form:
\begin{equation}
\label{diff_mat}
\frac{\diff \Psi}{\diff z} = \mathbf{C} \Psi.
\end{equation}

Since the permittivity is assumed to be independent of $z$ within a
slice of  the lens, the matrix $\mathbf{C}$ will be constant for
each slice.  Thus, Eq. (\ref{diff_mat}) can be integrated to obtain
\begin{equation}
\label{psi}
\Psi(z_b) = \exp\{(z_b-z_a)\mathbf{C}\}\Psi(z_a),
\end{equation}
for $z_a$ and $z_b$ inside the same slice. Let $z_j$ be at the
left-hand side of slice $j$, $\Delta_j$ the thickness of the
slice, and $\mathbf{C}_j$
 the $\mathbf{C}$-matrix for layer $j$. From  the field at $z_j$, we find the
field at the right hand side of the slice as
\begin{equation}
\Psi(z_j+\Delta_j) = \exp\{\Delta_j \mathbf{C}_j\}\Psi(z_j).
\end{equation}
Note that $j=0$ corresponds to the region between the source and
the lens, $j = 1,...,N$ are the slices inside the lens, and $j = N+1$
represents the region from the lens to
the image plane. The thicknesses are $\Delta_0 = a$, $\Delta_j = d/N$, for $j = 1,...,N$, and
$\Delta_{N+1} = b$. Let us  define
\begin{equation}
\label{Mj}
\mathcal{M}_j = \exp \{ \Delta_j \mathbf{C}_j\}.
\end{equation}
Eq. (\ref{Mj}) propagates the field from the start of a slice, to the end. Next, connecting the fields of adjacent slices with the electromagnetic boundary conditions, we find 
for the boundary between slice $j$ and $j+1$ that  
\begin{equation}
\frac{\partial 
    H_{j+1}(z_{j+1})}{\partial z} = \frac{\epsilon_{j+1}}{\epsilon_j} \frac{\partial
    H_j(z_{j+1})}{\partial z}.
\end{equation}
 Inserting the Fourier series from Eqs. (\ref{h-field}) and (\ref{permittivity}), we
 obtain a convolution on the right hand side corresponding
 to the multiplication of a Toeplitz matrix and the vector containing
the components $  \diff h_{m}(z)/\diff z$. The Toeplitz matrix
is defined by the Fourier components of $\epsilon_{j+1}/\epsilon_j$
  and will be called 
$\mathbf{P}$.  
 Then
 \begin{equation}
 \label{hz}
 \frac{\diff \mathbf{h}_j}{\diff z}(z_{j+1}) = \mathbf{P}_j(z_{j+1})
 \frac{\diff \mathbf{h}_{j+1}}{\diff z}(z_{j+1}),
 \end{equation}  
where $\mathbf{P}_j$ corresponds to the transition between layer $j$
and $j+1$.

Eq. (\ref{hz}) together with the fact that $\mathbf{h}$ is continuous across the layer boundary, gives us the following transfer
 matrix
 \begin{equation}
 \label{P}
 \left[
 \begin{array}[c]{c}
 \mathbf{h^+}_{j+1} \\ \mathbf{h^-}_{j+1}
 \end{array}
 \right] = \frac{1}{2} \left[
 \begin{array}[c]{cc}
  \mathbf{I} + \mathbf{k_z}^{-1} \mathbf{P}_j \mathbf{k_z}&
  \mathbf{I} - \mathbf{k_z}^{-1} \mathbf{P}_j \mathbf{k_z}\\
  \mathbf{I} - \mathbf{k_z}^{-1} \mathbf{P}_j \mathbf{k_z}&
  \mathbf{I} + \mathbf{k_z}^{-1} \mathbf{P}_j \mathbf{k_z}
 \end{array}
 \right] \left[
 \begin{array}[c]{c}
 \mathbf{h^+}_{j} \\ \mathbf{h^-}_{j}
 \end{array}
 \right].
 \end{equation} 
Let us call the transfer matrix in Eq. (\ref{P})  $\mathcal{P}_j$.

By the successive application of \eqref{Mj} and \eqref{P}, we find:
\begin{equation}
\label{eq:T}
 \left[
 \begin{array}
 [c]{c}
 \mathbf{T}\\
 \mathbf{0}
 \end{array}
 \right]
 = \mathcal{M}_{N+1} \prod_{j=N}^0 (\mathcal{P}_j \mathcal{M}_j)
 \left[
 \begin{array}
 [c]{c}
 \mathbf{I} \\
 \mathbf{R}
\end{array}
 \right].
 \end{equation}
Here, each column $i$ of $\mathbf{I}$ corresponds to an experiment where the
incident field amplitude is 1 for one of the Fourier components and
zero for the others. The $i$'th column of $\mathbf{R}$ is
the reflection at the source plane of experiment $i$, and the $i$'th
column of $\mathbf{T}$ is the corresponding transmission at the image
plane. To get the reflection in the case of two or more waves, the
corresponding columns of $\mathbf{R}$ are added, the new transmission
are found by adding colums of $\mathbf{T}$.
Once the total matrix in Eq. \eqref{eq:T} has been found, it is straightforward to calculate the
unknowns $\mathbf T$ and $\mathbf R$, and therefore the field amplitudes in all slices.

\section{Numerical results}
The thickness $d$ of the lens was chosen such that $\omega_0 d/c=2\pi/10$. The resolution clearly improves with decreasing distance $b$ from the lens to the image, since then the required evanescent fields at the end of the lens are reduced. However there may be practical reasons that makes it impossible to reduce the distances $a$ and $b$ below a certain value. In our simulations we have taken $a=b=d/2$. For simplicity we normalized $\epsilon_0=\mu_0=\omega_0=1$. The permeability was set to $\mu=-1$; however, since the lens was relatively thin, the specific value of $\mu$ did not matter significantly for evanescent TM waves. The number of slices were taken to be $N=20$. The computation domain $L$ was chosen in the range $(15,50)$ depending on the specific problem, and the number of Fourier components ($m$ values) was of the order of $100$.

First, a single mode source $\mathbf H = \exp(i k_x x+ i k_z z)\mathbf{\hat y}$ was considered. The reflection and transmission coefficients, and the fields in the lens, were computed using the iterative method above. The transmission coefficient is shown in Fig. \ref{fig:t}. It is easy to see improvements as a result of gain compensation, dependent on the saturation constant
$\mathcal{E}_\text{s}$. (Here it is useful to recall that for a normalized magnetic field $\mathbf H = \exp(i k_x x+ i k_z z)\mathbf{\hat y}$, the electric field is $\mathbf E=([k_z/\omega_0\epsilon\epsilon_0]\mathbf{\hat x} - [k_x/\omega_0\epsilon\epsilon_0]\mathbf{\hat z})\exp(i k_x x+ i k_z z)$. Thus, for our normalization, we see that $|\mathbf E|\sim k_x$, to be compared to $\mathcal E_s$.) Nevertheless, for a fixed amplitude of the incident field, Fig. \ref{fig:res} indicates that an exponential increase in the saturation constant is needed for a linear increase in the resolution. This is an important result, as it shows the difficulty of getting large resolution: The required, large evanescent fields associated with large spatial frequencies saturate the gain at the output of the lens.

The reflection coefficient is plotted in Fig. \ref{fig:r}.
We note that significant reflections arise even for the spatial frequencies where the transmission
is relatively large. As can be seen in Fig. \ref{fig:efelt}, the field distributions of the two
evanescent components in the lens increase roughly exponentially with $+z$ or $-z$, respectively.
For small spatial frequencies, where the lens is essentially perfect, the one increasing in the $+z$
direction dominates. For higher spatial frequencies the two components have a similar amplitude,
such that the total field and therefore the imaginary part of the permittivity start to look like a
U-shaped valley.

In general, different plane wave components of the source will couple to each other through Eq.
\eqref{eq:chia0}. To simulate the gain-compensated lens under more real-world conditions, it was therefore tested with
several waves traversing the lens simultanously.
The transmission of one wave as a function of $k_x$, in the presence
of another wave $-k_x$, is shown in Fig. \ref{fig:t}.
The amplitudes of both waves were set to $1/2$ to keep the
total field at the source equal to the case with a single wave. Moreover, from a number of simulations with
several waves, a useful rule of thumb was discovered: As a worst-case estimate, one can judge whether the lens
operates as required by assuming that the mode with largest $k_x$ has amplitude equal to the sum of
the amplitudes at the source. More preciesly, suppose a single mode $k_x$ with amplitude
$1$ experiences a transmission greater than $1/2$. 
For any superposition of modes with transversal wavenumbers less than $k_x$ and sum of amplitudes equal to unity, each 
mode will experience a transmission greater than $1/2$.

For conventional lenses, the Rayleigh criterion is usually applied to
quantify the distance between two point sources (or in the
one-dimensional case, line sources) in order to resolve their images.
Since our lens is nonlinear, the image of two line sources cannot be
determined as a superposition of the fields associated with the two
sources separately, or as a superposition of the fields associated with
their Fourier components. Therefore, as in previous literature on
perfect lenses, we have chosen to consider a single Fourier component
source, and defined the spatial resolution as $2\pi/k_x$, where $k_x$ is
the half maximum wavenumber (Fig. \ref{fig:res}). Note that the behavior
of the lens for more complex sources can be determined from the rule of
thumb described above.

Fig. \ref{fig:slits} shows the absolute value of the transmitted magnetic field at the image plane, $|H(x,2d)|$, resulting 
from a source consisting of two slits. The image of the slits are
clearly better resolved with an increased saturation constant $\mathcal{E}_s$.

\section{Conclusion}
We have developed a method for calculating the transmission, reflection, and detailed field profile
of a gain-compensated perfect lens, taking into account gain saturation. The gain compensation
clearly improves the resolution limit of perfect lenses. However, due to gain saturation, a number
of nonideal effects arise, included limited resolution and reflections. The nonideal effects depend
heavily on the saturation constant and/or the field strength of the source. 

If there are different waves traversing the lens at the same time, they will interact through the
material. Waves with a spatial frequency close to the resolution limit will have the greatest
impact. As a rule of thumb, it is enough to know the sum of amplitudes of the waves at the
source, and then assume the mode with the largest spatial frequency has this amplitude.
If this single wave is transmitted, in the sense of a transmission larger than $1/2$, then so will
any superposition of waves with less spatial frequencies and the same sum of amplitudes.

The calculations in this work was performed for TM polarization and a one-dimensional source. For a
two-dimensional source with both polarizations, both dielectric and magnetic losses should be
compensated, that is, $\im\epsilon$ and $\im\mu$ must be reduced. Although the theory in this paper
can trivially be extended to this situation, there may be serious practical problems associated with
the fabrication of such active media for optical frequencies.

For a noncompensated lens, the maximum spatial frequency resolved by the lens is approximately
$-(1/d)\ln(|1+\epsilon|/2)$ \cite{ramakrishna2002,johansen}. Thus, for a fixed $d$, an
exponential decrease in the losses is necessary to increase the resolution linearly. From our
numerical results, a similar relation is approximately valid for the
saturation constant of a gain compensated medium; to achieve
a linear improvement in the resolution, the saturation constant must increase exponentially. This
clearly shows the difficulties of achieving very high resolution.

%\section{References}

\clearpage

\section*{List of Figure Captions}

Fig. 1. (Color online) Perfect lens in vacuum. The parameter $d$ is the thickness of the lens, $a$ and $b$ are the
distances from the source to the lens, and from the lens to the image
plane, respectively. The parameters are
governed by the equation $d = a + b$. The numbers $1$ through $N$ indicate the different slices. The
lens is considered to be infinite in the $xy$-plane.

Fig. 2. (Color online) The absolute value of the transmission coefficient  when $\omega_0/c=1$ (normalized), $\omega_0 d/c=2\pi/10$, $a=b=d/2$, $\im\chi_\text{p}(\omega_0)=0.05$, and $N=20$:
($a$) Non-compensated lens;
($b$) $\mathcal{E}_\text{s}=10$, $\Delta\chi=0$;
($c$) $\mathcal{E}_\text{s}=4$, $\Delta\chi=0$; 
($d$) $\mathcal{E}_\text{s}=10$, $\Delta\chi=0.015$;
($e$) $\mathcal{E}_\text{s}=10$, $\Delta\chi=0$, two waves, $k_x$ and $-k_x$,
both having
amplitude $1/2$.

Fig. 3. (Color online) The resolution of the lens as a function of the saturation constant. The resolution is
defined as the $k_x$-value where the transmission equals $1/2$.
Parameters: $\omega_0/c=1$, $\omega_0 d/c=2\pi/10$, $\im\chi_p(\omega_0)=0.05$, $\Delta\chi=0$,
$a=b=d/2$, and $N=20$.

\begin{figure}[t!]
\includegraphics[width=0.95\textwidth]{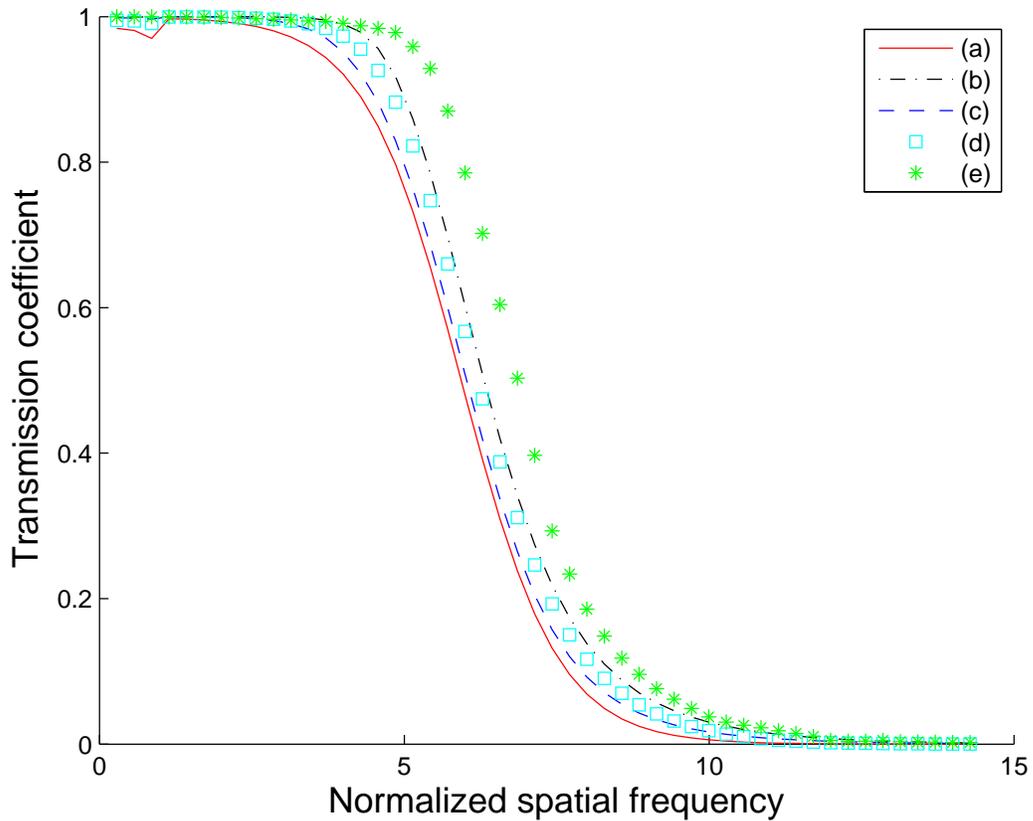}
\caption{(Color online) The absolute value of the transmission coefficient  when $\omega_0/c=1$ (normalized), $\omega_0 d/c=2\pi/10$, $a=b=d/2$, $\im\chi_\text{p}(\omega_0)=0.05$, and $N=20$:
($a$) Non-compensated lens;
($b$) $\mathcal{E}_\text{s}=10$, $\Delta\chi=0$;
($c$) $\mathcal{E}_\text{s}=4$, $\Delta\chi=0$; 
($d$) $\mathcal{E}_\text{s}=10$, $\Delta\chi=0.015$;
($e$) $\mathcal{E}_\text{s}=10$, $\Delta\chi=0$, two waves, $k_x$ and $-k_x$,
both having
amplitude $1/2$. ASHKSF2.eps.
}
\label{fig:t}
\end{figure}

\begin{figure}[t!]
\includegraphics[width=0.95\textwidth,viewport=100 240 550 560]{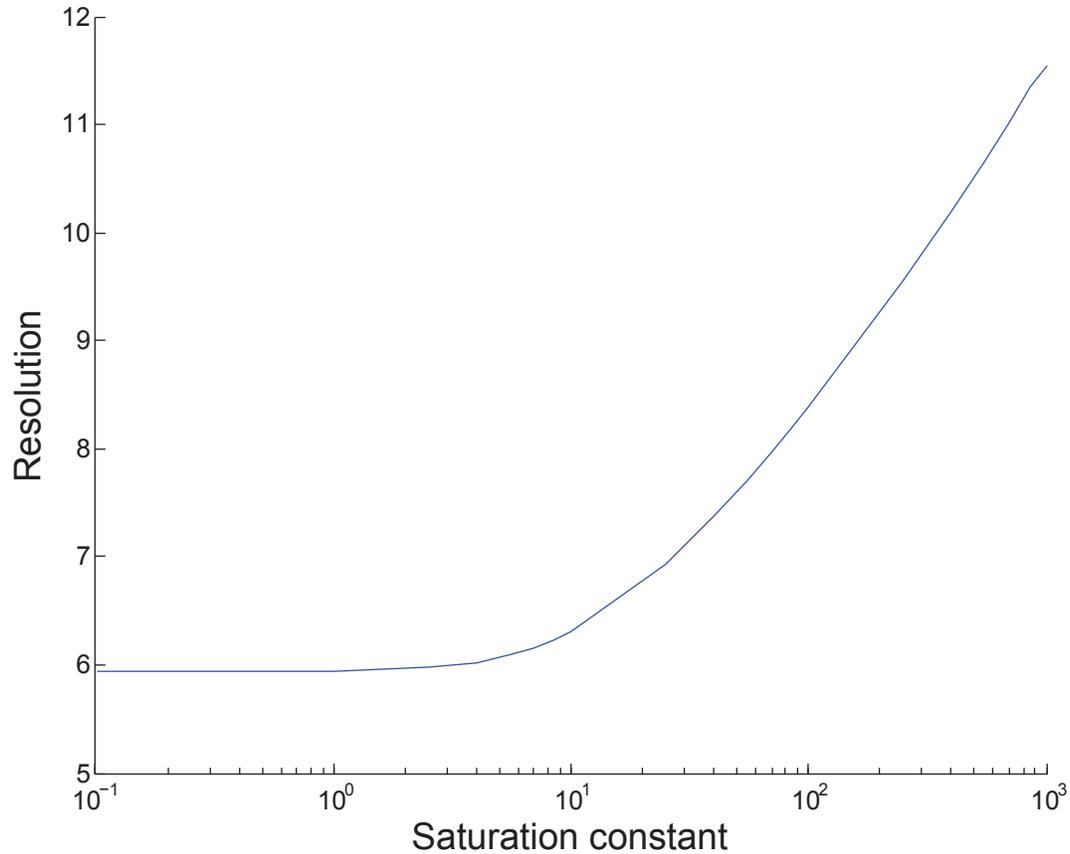}

\caption{(Color online) The resolution of the lens as a function of the saturation constant. The resolution is
defined as the $k_x$-value where the transmission equals $1/2$.
Parameters: $\omega_0/c=1$, $\omega_0 d/c=2\pi/10$, $\im\chi_p(\omega_0)=0.05$, $\Delta\chi=0$,
$a=b=d/2$, and $N=20$. ASHKSF3.eps. 
}
\label{fig:res}
\end{figure}

\begin{figure}[htpb]
\includegraphics[width=0.95\textwidth]{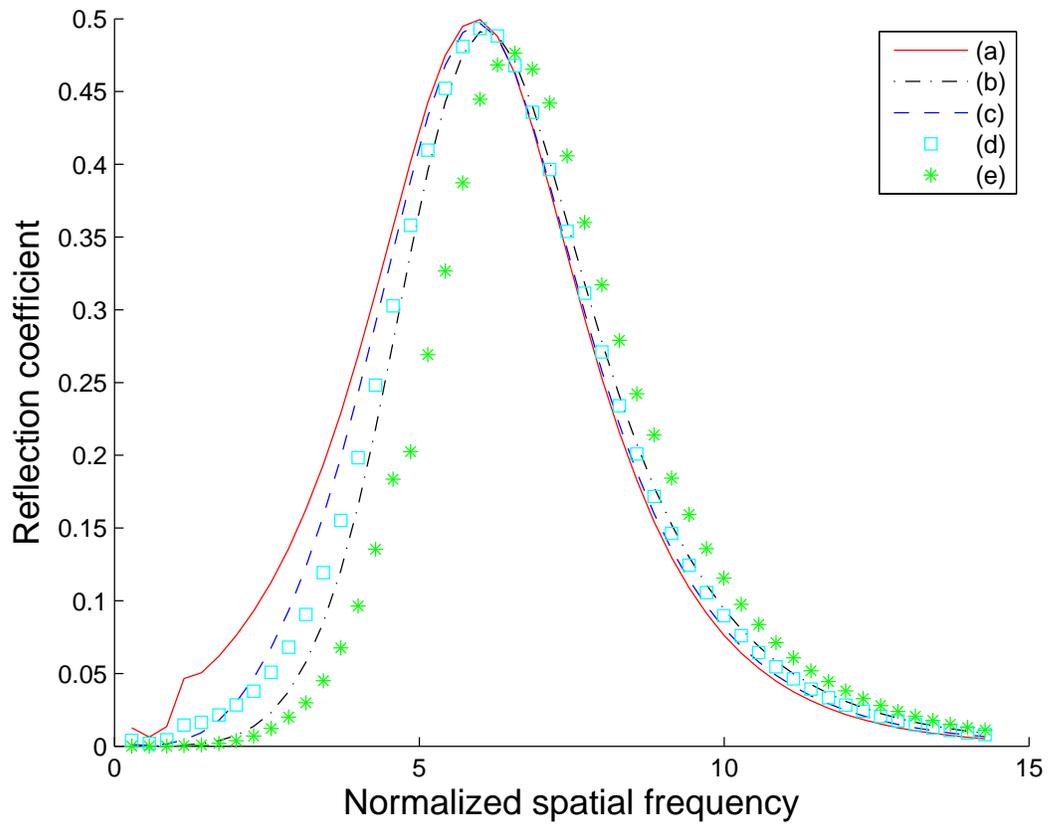}
\caption{(Color online) The absolute value of the reflection coefficient at the
  source plane after convergence, for the same cases as those in
Fig. \ref{fig:t}. ASHKSF4.eps.
}
\label{fig:r}
\end{figure}

\begin{figure}[htpb]
\includegraphics[width=0.95\textwidth,viewport=100 240 550 560]{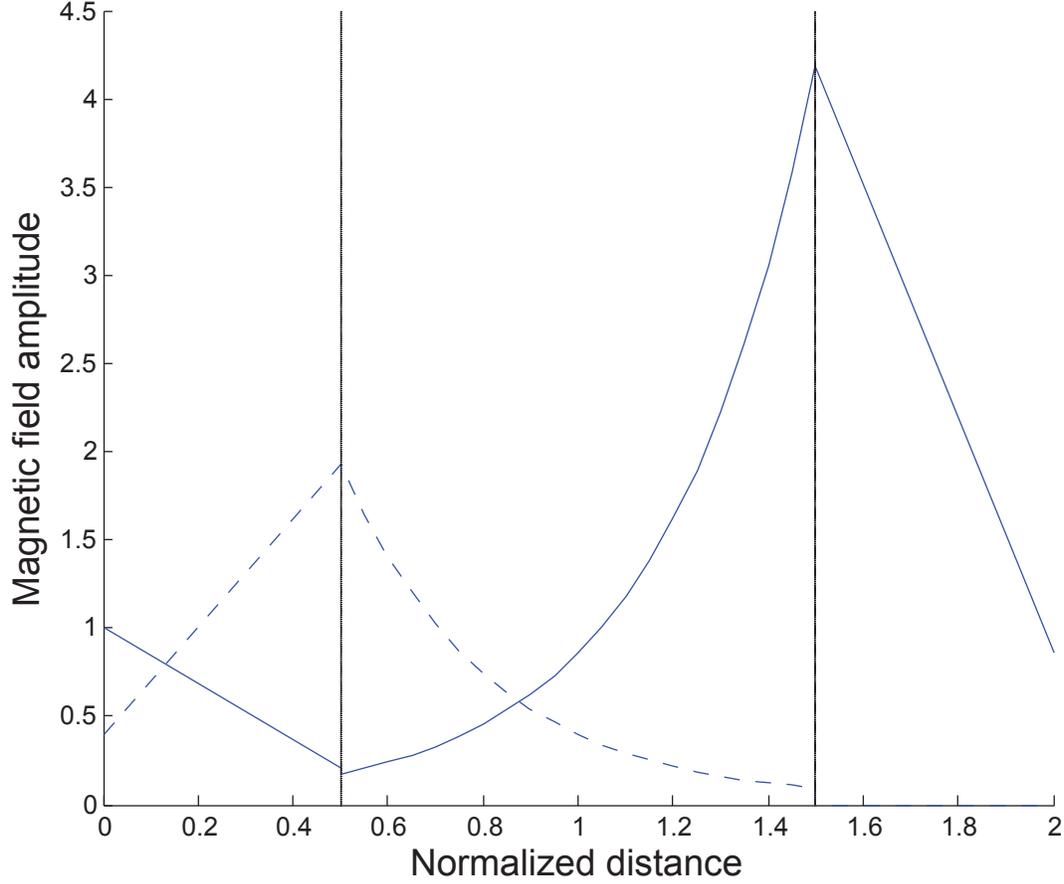}
\caption{
(Color online) The distribution of the two components of the evanescent field in the
lens, for one wave with $k_x=5.1408$. The distance is normalized with respect to
lens thickness, $d$. 
Parameters: $\omega_0/c=1$, $\omega_0 d/c=2\pi/10$, $\im\chi_p(\omega_0)=0.05$, $\Delta\chi=0$,
$\mathcal{E}_\text{s}=10$, $a=b=d/2$, and $N=20$. The solid line shows the
absolute value of the nonzero component of $\mathbf{h}^+$, and the
dotted line shows the
absolute value of the nonzero component of $\mathbf{h}^-$. ASHKSF5.eps.
}
\label{fig:efelt}
\end{figure}

\begin{figure}[htpb]
\includegraphics[width=0.95\textwidth]{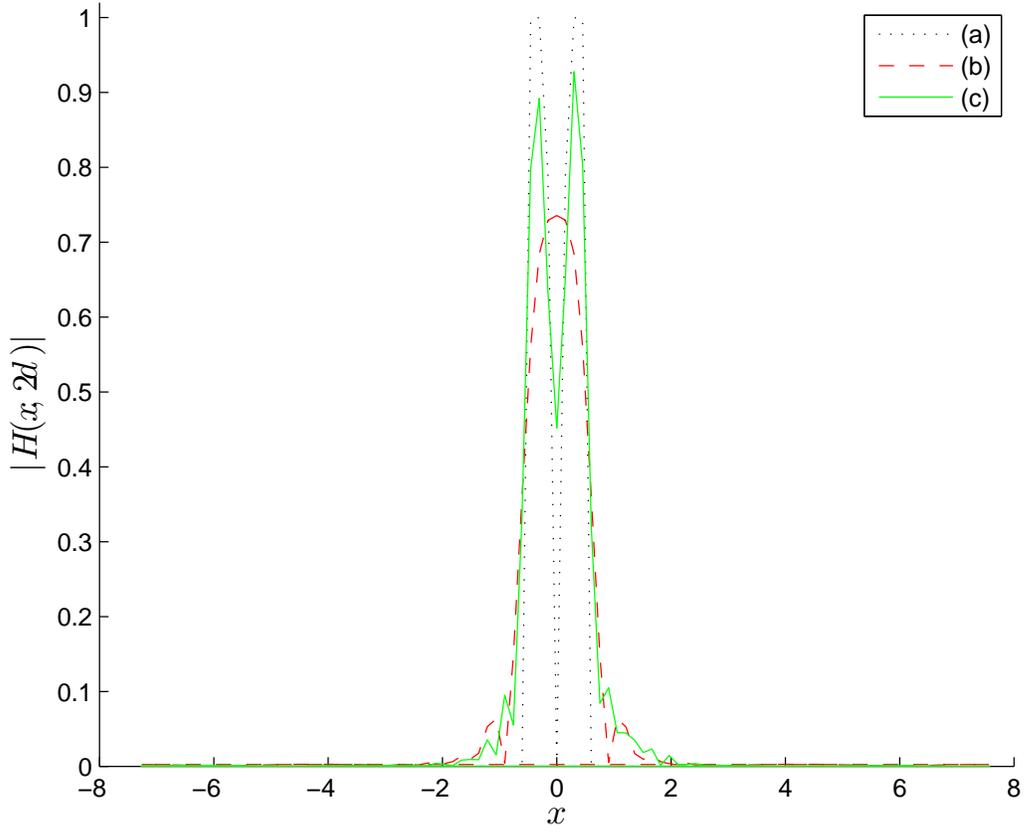}
\caption{
(Color online) The absolute value of the transmitted magnetic field at the image plane, when the source consist of two
slits. Parameters:
$\omega_0/c=1$, $\omega_0 d/c=2\pi/10$, $\im\chi_p(\omega_0)=0.05$,
$\Delta\chi=0$, $a=b=d/2$, and $N=20$: (a) The incident magnetic field
at the source, (b)
$\mathcal{E}_s = 0.1$, and (c) $\mathcal{E}_s= 50$. ASHKSF6.eps.
}
\label{fig:slits}
\end{figure}

Fig. 4. (Color online) The absolute value of the reflection coefficient at the
  source plane after convergence, for the same cases as those in
Fig. \ref{fig:t}.

Fig. 5. (Color online) The distribution of the two components of the evanescent field in the
lens, for one wave with $k_x=5.1408$. The distance is normalized with respect to
lens thickness, $d$. 
Parameters: $\omega_0/c=1$, $\omega_0 d/c=2\pi/10$, $\im\chi_p(\omega_0)=0.05$, $\Delta\chi=0$,
$\mathcal{E}_\text{s}=10$, $a=b=d/2$, and $N=20$. The solid line shows the
absolute value of the nonzero component of $\mathbf{h}^+$, and the
dotted line shows the
absolute value of the nonzero component of $\mathbf{h}^-$.

Fig. 6. (Color online) The absolute value of the transmitted magnetic field at the image plane, when the source consist of two
slits. Parameters:
$\omega_0/c=1$, $\omega_0 d/c=2\pi/10$, $\im\chi_p(\omega_0)=0.05$,
$\Delta\chi=0$, $a=b=d/2$, and $N=20$: (a) The incident magnetic field
at the source, (b)
$\mathcal{E}_s = 0.1$, and (c) $\mathcal{E}_s= 50$.

\clearpage

\end{document}